\documentclass[12pt]{article}
\usepackage{amsmath,amssymb,mathrsfs,url,hyperref}

\title{Detection Time Distribution for Several Quantum Particles}

\author{
Roderich Tumulka\footnote{Fachbereich Mathematik,
	Eberhard-Karls-Universit\"at, Auf der Morgenstelle 10,
	72076 T\"ubingen, Germany.
	E-mail: roderich.tumulka@uni-tuebingen.de}
}

\date{October 8, 2022}

\newcommand{\Hilbert}{\mathscr{H}}
\newcommand{\sS}{\mathscr{S}}
\newcommand{\be}{\begin{equation}}
\newcommand{\ee}{\end{equation}}
\newcommand{\scp}[2]{\langle #1|#2\rangle}

\renewcommand{\Im}{\mathrm{Im}}
\newcommand{\RRR}{\mathbb{R}}
\newcommand{\CCC}{\mathbb{C}}

\newcommand{\vx}{\boldsymbol{x}}

\newcommand{\vn}{\boldsymbol{n}}

\newcommand{\vX}{{\boldsymbol{X}}}
\newcommand{\vj}{{\boldsymbol{j}}}
\newcommand{\vu}{{\boldsymbol{u}}}
\newcommand{\vv}{{\boldsymbol{v}}}

\newcommand{\vsigma}{{\boldsymbol{\sigma}}}
\newcommand{\Q}{\mathcal{Q}}
\newcommand{\Lab}{\mathscr{I}}
\newcommand{\R}{\Omega}
\newcommand{\bou}{\partial \R}
\newcommand{\bouS}{\partial_+ S}
\newcommand{\bouSt}{\partial_+ \tilde{S}}
\newcommand{\prob}{\mathrm{Prob}}
\newcommand{\Z}{\mathscr{Z}}

\begin{document}
\maketitle
\begin{abstract}
We address the question of how to compute the probability distribution of the time at which a detector clicks, in the situation of $n$ non-relativistic quantum particles in a volume $\R\subset \RRR^3$ in physical space and detectors placed along the boundary $\bou$ of $\R$. We have previously \cite{detect-rule} argued in favor of a rule for the 1-particle case that involves a Schr\"odinger equation with an absorbing boundary condition on $\bou$ introduced by Werner; we call this rule the ``absorbing boundary rule.'' Here, we describe the natural extension of the absorbing boundary rule to the $n$-particle case. A key element of this extension is that, upon a detection event, the wave function gets collapsed by inserting the detected position, at the time of detection, into the wave function, thus yielding a wave function of $n-1$ particles. We also describe an extension of the absorbing boundary rule to the case of moving detectors.

\medskip

\end{abstract}

\section{Introduction}

The question of how to compute the probability distribution of the time at which a detector registers a quantum particle from the initial wave function $\psi_0$ of that particle has attracted considerable interest in the literature; see, e.g., \cite{AB61,All69,Kij74,GRT96,AOPRU98,ML00,Bau00,HSM03,MSE08,MRC09,VHD13,Dhar13} and the references therein. However, the most convincing equations for this, written down in 1987 by Werner \cite{Wer87}, have received little if any attention. We have proposed \cite{detect-rule} to take these equations as a rule for computing the detection time distribution of an ideal detector (and thus as  the definition of an ``ideal detector''); we call this rule the \emph{absorbing boundary rule}. We have pointed out in \cite{detect-rule} why it achieves exactly what one should hope for in such a rule and given a novel derivation of it. In the present paper, we develop an extension of the absorbing boundary rule to the case of $n$ particles able to trigger detectors, as well as an extension to the case of moving detectors (resulting in a Galilean-covariant theory). 

A brief remark about the name: The absorbing boundary rule involves a boundary condition that we call the \emph{absorbing boundary condition}. These names do not mean that a wave reaching the boundary will be completely absorbed; in fact, a significant part of it will be reflected \cite[Sec.~3]{detect-rule}. The names mean only that \emph{some} part of the wave gets absorbed, in contrast to the well-known boundary conditions of Dirichlet and Neumann. (And in the Bohmian picture, the particle gets absorbed with certainty when reaching the boundary.) Our goal here is not to model an ideal absorber, but an ideal detector. 

Absorbing boundaries have also been considered in numerical analysis for the purpose of solving the Schr\"odinger equation in unbounded regions while avoiding any unphysical reflection of the wave from the boundary of the numerical grid \cite{FJ99,SK10,Sel21}. However, for this aim it is preferable to use methods that absorb more of the wave than the boundary condition considered here; but also there the question how to treat several particles arises (see also Remark~\ref{rem:several-imaginary} in Section~\ref{sec:discussion}). 

We present an analogous rule for the Dirac equation (for one or more particles) in \cite{detect-dirac}; an alternative derivation of the 1-particle rule through a suitable limit of repeated position measurements is described in \cite{DBD20}. 

Suppose that $n$ particles are, at the initial time $t_0$, located in a region $\R\subset \RRR^3$ in physical space with wave function $\psi_0\in L^2(\R^n)$, and that detectors are placed along the boundary $\bou$ of $\R$. For each particle $i\in\{1,\ldots,n\}$, we consider the time $T_i$ and the location $\vX_i\in\bou$ where it gets detected and write $Z_i=(T_i,\vX_i)$; should it never be detected, we set $Z_i=\infty$. The $n$-particle rule proposed in this paper specifies the joint probability distribution of $Z_1,\ldots,Z_n$ in terms of $\psi_0$. It is of the form
\be\label{povm}
\prob_{\psi_0}\Bigl\{ (Z_1,\ldots,Z_n)\in B \Bigr\} = \scp{\psi_0}{E(B)|\psi_0}
\ee
for all sets $B\subseteq \bigl( [t_0,\infty)\times \bou \cup\{\infty\} \bigr)^n$ with a certain positive-operator-valued measure (POVM) $E(\cdot)$. Like the rule for a single particle, it provides an example of an observable that is not given by a self-adjoint operator (for which the $E(B)$ would be projections) but a proper POVM instead.

The status of the proposed $n$-particle rule is partly that of a definition of a concept of ``ideal detector,'' and partly the upshot of a theoretical analysis of this kind of experiment. Although there is no self-adjoint time operator, quantum mechanics makes a prediction for this experiment, by treating the detectors as an $N$-particle quantum system, solving the Schr\"odinger equation for the whole system of $N+n$ particles, and computing from that wave function the probability at any given $t\geq t_0$ that the detector at $\vx\in\bou$ is in a triggered state. Since this method of obtaining a prediction is impractical, it is desirable to make idealizing assumptions, and any rule for the joint distribution of $Z_1,\ldots,Z_n$ in terms of $\psi_0$ would involve some such assumptions. We describe the reasoning and assumptions that lead to the proposed rule in Section~\ref{sec:derivation}.

This paper is organized as follows. In Section~\ref{sec:rule}, we describe the absorbing boundary rule for a single particle with fixed detectors as in \cite{detect-rule}. In Section~\ref{sec:moving}, we describe our extension to moving detectors. In Section~\ref{sec:several}, we describe our extension to $n$ particles, and in Section~\ref{sec:derivation} we provide a derivation. Finally, in Section~\ref{sec:discussion}, we provide further remarks and discussion.

\section{Statement of the Rule for a Single Particle}
\label{sec:rule}

Consider a single non-relativistic particle of mass $m>0$. According to the absorbing boundary rule, the particle's wave function $\psi$ evolves according to the Schr\"odinger equation
\be\label{Schr}
i\hbar\frac{\partial \psi}{\partial t} = -\frac{\hbar^2}{2m} \nabla^2 \psi + V\psi
\ee
in $\R$ with potential $V:\R\to\RRR$ and boundary condition
\be\label{bc}
\frac{\partial \psi}{\partial n}(\vx) = i\kappa \psi(\vx)
\ee
at every $\vx\in\bou$, where $\kappa>0$ is a constant characterizing the type of detector (the wave number of sensitivity) and $\partial/\partial n$ denotes the outward normal derivative on the surface $\bou$, i.e.,
$\partial \psi/\partial n := \vn(\vx) \cdot \nabla\psi(\vx)$
with $\vn(\vx)$ the outward unit normal vector to $\bou$ at $\vx\in\bou$. It is known \cite{Wer87,detect-thm} that \eqref{Schr} and \eqref{bc} together have a unique solution $\psi_t$ for every $\psi_0\in L^2(\R)$. Assuming that $\|\psi_0\|^2 = \int_\R d^3\vx\, |\psi_0(\vx)|^2=1$, the rule asserts that the distribution of $Z=(T,\vX)$, where $T$ is the exit time and $\vX$ the exit location on $\bou$,
\begin{equation}\label{probnjR}
  \prob_{\psi_0} \Bigl( t_1 \leq T<t_2, \vX \in B \Bigr) =
  \int\limits_{t_1}^{t_2} dt \int\limits_{B} d^2\vx \; \vn(\vx) \cdot
  \vj^{\psi_t}(\vx)
\end{equation}
for any $t_0\leq t_1<t_2$ and any set $B\subseteq \bou$, with $d^2\vx$ the surface area element and $\vj^\psi$ the probability current vector field
defined by $\psi$, which is
\begin{equation}\label{jSchr}
  \vj^\psi = \frac{\hbar}{m} \Im\, \psi^* \nabla \psi\,.
\end{equation}
(By \eqref{bc}, the integrand in \eqref{probnjR} can also be written as $\frac{\hbar\kappa}{m}|\psi_t(\vx)|^2$.) Finally,
\begin{align}\label{Zinfty}
\prob_{\psi_0}(Z=\infty) 
&= 1-  \int\limits_{t_0}^{\infty} dt \int\limits_{\bou} d^2\vx \; \vn(\vx) \cdot
  \vj^{\psi_t}(\vx)\\
&=\lim_{t\to\infty} \int\limits_{\R}d^3\vx\; |\psi_t(\vx)|^2\,.
\end{align}

It is easy to verify \cite{detect-rule} that \eqref{probnjR} and \eqref{Zinfty} define a probability distribution on $\Z=[t_0,\infty)\times \bou\cup \{\infty\}$, the value space of $Z$. 

The distribution \eqref{probnjR} can equivalently be characterized as the probability distribution of the space-time location where a Bohmian trajectory $\vX(t)$ \cite{DT}, starting at time $t_0$ from a $|\psi_0|^2$-distributed random initial position in $\R$, will cross $\bou$ if guided, according to Bohm's equation of motion $d\vX/dt= \vj^{\psi_t}(\vX)/|\psi_t(\vX)|^2$, by $\psi_t$ obeying \eqref{Schr} and \eqref{bc} (and $Z=\infty$ if the Bohmian particle never reaches $\bou$) \cite{Daumer,detect-rule}. In fact, in the Bohmian picture a detector clicks when and where the particle arrives, but the presence of detectors has an influence on the particle's motion even before any detection, as \eqref{bc} would have to be dropped in the absence of detectors. The boundary condition \eqref{bc} enforces that the particle can cross the detecting surface $\bou$ only outward, in fact at a normal velocity of $\hbar\kappa/m$.

The distribution \eqref{probnjR} corresponds to a POVM $E_\kappa(\cdot)$ on $\Z$ that can be expressed as
\begin{align}
  E_\kappa \bigl( dt\times d^2\vx \bigr) 
  &= \frac{\hbar\kappa}{m} \, W_{t-t_0}^\dagger \,
  |\vx\rangle\langle\vx|  \,W_{t-t_0}\,  dt\, d^2\vx\\
E_\kappa (\{\infty\}) 
&= I-E_\kappa([t_0,\infty)\times \bou)
=\lim_{s\to\infty} W_s^\dagger W_s 
\end{align}
with ${}^\dagger$ denoting the adjoint operator and $W_s$ the (non-unitary) linear operator that maps $\psi_{t_0}=\psi_0$ to $\psi_{t_0+s}$ solving \eqref{Schr} and \eqref{bc}. The operators $W_s$ for $s\geq 0$ have the properties $W_0=I$, $W_t W_s=W_{t+s}$, and $\|W_s\psi\|\leq \|\psi\|$; that is, they form a \emph{contraction semigroup}. The quantity $\|\psi_t\|^2$ equals the probability that $Z\geq t$. Werner \cite{Wer87} based his terminology of exit time and exit location on an \emph{arbitrary} contraction semigroup and presented the boundary condition \eqref{bc} only as an example; later on, he turned to other approaches concerning the detection time distribution \cite{Wer88}.

The absorbing boundary rule describes detectors that get triggered immediately when the particle reaches the surface $\bou$---a concept that might have seemed impossible in view of the quantum Zeno effect \cite{Fri72,Zeno,Dhar13,Dhar14}. The rule also makes clear which wave function to attribute to the particle as the collapsed wave function, should the experiment (i.e., the attempted detection) be terminated without detection event at any time $t\geq t_0$; that collapsed wave function is $\psi_t/\|\psi_t\|$. It is different from what the wave function would have been in the absence of detectors, i.e., from the solution of the Schr\"odinger equation \eqref{Schr} in $\RRR^3$ without a boundary at $\bou$, and without boundary condition \eqref{bc}.

\section{Extension to Moving Boundaries}
\label{sec:moving}

We now allow the region $\R=\R(t)$ to be (smoothly) time-dependent, which corresponds
to moving detectors and defines a space-time region $S = \{(t,\vx):
t\geq t_0, \vx \in \R(t)\}$ available to the particle. In this case, the
value space of the random variable $Z$ is $\Z = \bouS \cup
\{\infty\}$, where $\bouS$ is the set of boundary points $(t,\vx)$ of $S$ with $t>t_0$ (since $(t,\vx)$ with $t=t_0$ and $\vx\in\R(t_0)$ are not exit points). We propose the following variant of the absorbing boundary rule.

The wave function satisfies the Schr\"odinger equation \eqref{Schr} inside $S$ with the boundary condition \eqref{bc} replaced by
\be\label{bct}
\vn(t,\vx)\cdot \nabla \psi(t,\vx) = i\kappa(t,\vx)\, \psi(t,\vx)
\ee
for $\vx\in\bou(t)$, where $\vn(t,\vx)$ is the outward-pointing unit normal vector to $\bou(t)$ at $\vx\in\bou(t)$. The admissible values of $\kappa(t,\vx)$ are constrained by the condition
\be\label{constraint}
\frac{\hbar\kappa(t,\vx)}{m} \geq v_n(t,\vx)\,,
\ee
where $v_n$ is the (possibly negative) normal velocity at which the boundary is moving outward.
The constraint \eqref{constraint} can be understood in Bohmian mechanics as ensuring that all Bohmian trajectories on the boundary are crossing the boundary in the outward direction; since it forces the normal outward velocity to be $\hbar\kappa(t,\vx)/m$, this value must be greater than the speed at which the boundary is moving outward, or else the trajectory cannot reach the boundary (and, instead, a trajectory is beginning at this boundary point and continuing inside $\R$). The condition \eqref{constraint} replaces the condition $\kappa\geq0$, and indeed $\kappa(t,\vx)$ can be negative when $v_n(t,\vx)$ is, as long as \eqref{constraint} is satisfied. The appropriate value of $\kappa(t,\vx)$ when all detectors have wave number of sensitivity $\kappa$ 
is
\be
\kappa(t,\vx) = \kappa + \frac{mv_n(t,\vx)}{\hbar}\,.
\ee

The probability distribution of $(T,\vX)$ is then exactly the distribution of where the Bohmian trajectory, starting at time $t_0$ at a location $\vX(t_0)$ that is $|\psi_0|^2$-distributed, hits $\bouS$. 
Explicitly, for any $B\subseteq \bouS$,
\begin{align}
\prob(Z\in B) 
&= \int_B dt \, d^2\vx\, \Bigl(\vn(t,\vx) \cdot \vj^{\psi_t}(\vx) - v_n(t,\vx)\, |\psi_t(\vx)|^2\Bigr) \\
&= \int_B dt \, d^2\vx\, |\psi_t(\vx)|^2\,\biggl( \frac{\hbar\kappa(t,\vx)}{m} - v_n(t,\vx)\biggr) \,.
\end{align}
Note that the integrand is non-negative by virtue of \eqref{constraint}. Furthermore, again, $\prob(Z=\infty)= \lim_{t\to\infty}\|\psi_t\|^2$.

This theory of a moving detecting surface is invariant under Galilean transformations. Indeed, under a Galilean boost with relative velocity $\vv$, every space-time point $(t,\vx)$ gets new coordinates $(t,\vx+\vv t)$, and a 1-particle wave function transforms according to
\be
\tilde{\psi}_t(\vx) = \exp\Bigl(\frac{im}{\hbar}  (\vv \cdot \vx - \tfrac{1}{2} v^2 t)\Bigr) \, \psi_t(\vx-\vv t)\,.
\ee
The detecting surface transforms to $\bouSt=\{(t,\vx+\vv t): (t,\vx)\in \bouS\}$, which still has the same normal vectors in space, $\tilde{\vn}(t,\vx+\vv t) = \vn(t,\vx)$, but changed normal velocity $\tilde{v}_n(t,\vx+\vv t) = v_n(t,\vx) + \vv\cdot \vn(t,\vx)$; and $\kappa$ in \eqref{bct} gets replaced by $\tilde{\kappa}(t,\vx+\vv t) = \kappa(t,\vx) + m \vv \cdot \vn(t,\vx)/\hbar$ (as expected in view of the Bohmian trajectories).

\section{Proposed Rule for Several Particles}
\label{sec:several} 

Now consider $n$ particles that can be detected, and let $\Lab$ be an index set of size $n$ containing the labels of the particles (e.g., $\Lab=\{1,\ldots,n\}$). We write $x=(\vx_i)_{i\in\Lab}$ for a configuration of the $n$ particles. (The symbol $i$ can either denote a particle label or $\sqrt{-1}$; it should always be clear which one is meant.) For simplicity, we return to a scenario of non-moving detectors; the general case of $n$ particles and moving detectors is described in Remark~\ref{rem:moving-several} in Section~\ref{sec:discussion}. We can allow the greater generality of having a separate detecting surface for each particle $i$; so let $\R_i\subseteq \RRR^3$ be the region in which particle $i$ is located initially, and suppose that detectors sensitive to particle $i$ are located along $\bou_i$.\footnote{This includes the possibility that $\R_i=\RRR^3$ for some particles for which no detectors are present. It is also straightforward to deal with the possibility that some particles (for which suitable detectors may or may not be present) get reflected, rather than absorbed, at certain surfaces; this would correspond to imposing, instead of \eqref{bc} or \eqref{bcn}, a boundary condition implying $\vn\cdot\vj=0$, such as a homogeneous Dirichlet or Neumann condition, on some boundary surfaces.} We take for granted that the initial wave function $\psi_0(x)$ is supported in the region $\R:= \prod_{i\in\Lab} \R_i$ of $3n$-dimensional configuration space. We will denote by $T^j$ and $\vX^j$ the time and location of the $j$-th detection event, while $T_i$ and $\vX_i$ (with lower indices) denote the time and location of the detection of particle $i$; so $0\leq T^1\leq T^2\leq \ldots$. We write $I^j$ for the label for the $j$-th detected particle and $Z^j=(T^j,I^j,\vX^j)$. We take for granted that a detected particle gets absorbed (or removed from the system; for another possibility see Remark~\ref{rem:outside} in Section~\ref{sec:discussion}). Thus, after the first detection event, 
the other $n-1$ particles continue moving around, and the detectors wait for any of them to arrive. 
If fewer than $n$ clicks ever occur, say only $r<n$, then we set $Z^j=\infty$ for $j>r$. Writing $Z_i$  (with lower index) for the detection time and location of the $i$-th particle, the $Z^1,\ldots,Z^n$ are more or less the re-ordering of the $Z_1,\ldots,Z_n$ (if $\Lab=\{1,\ldots,n\}$) in the order of detection. Our rule will specify the joint distribution of $Z=(Z^1,\ldots,Z^n)$ in $\Z= \bigl([t_0,\infty)\times\Lab\times(\cup_i \bou_i)\cup\{\infty\}\bigr)^n$. 
The boundary $\bou$ of $\R$ in $\RRR^{3n}$ is the union of $n$ faces,
\be
 F_i =  \bou_i\times \prod_{k\in \Lab\setminus\{i\}} \R_k 
\ee
for $i\in\Lab$,
together with a set of lower dimension (edges) that can be ignored because the probability current into that set vanishes. Put differently, the configuration reaches the boundary of $\R$ exactly when, for one $i\in\Lab$, particle $i$ reaches $\bou_i$.

\subsection{Statement of the $n$-Particle Rule}

The wave function evolves according to the Schr\"odinger equation
\be\label{Schrn}
i\hbar\frac{\partial \psi}{\partial t} = -\sum_{i\in\Lab} \frac{\hbar^2}{2m_i} \nabla_i^2 \psi + V\psi
\ee
in $\R\subseteq \RRR^{3n}$ with the boundary condition
\be\label{bcn}
\vn_i(\vx_i)\cdot \nabla_i\psi(x) = i\kappa_i \psi(x)
\ee
for all $x\in F_i$ and $i\in\Lab$, with positive constants $\kappa_i$. We assume that the potential is of the form
\be\label{V}
V(x) = \sum_{i\in\Lab} V_i(\vx_i) + \frac{1}{2}\sum_{\substack{i,j\in\Lab\\i\neq j}}V_{ij}(\vx_i,\vx_j)\,.
\ee
As in the 1-particle case, the probability current at the boundary $\bou$ is always outward-pointing. We postulate that, assuming $\|\psi_0\|^2=\int_\R d^{3n}x\,|\psi_0(x)|^2=1$, the joint distribution of $T^1,I^1,\vX^1$ is given by
\begin{multline}\label{probnjRn}
  \prob_{\psi_0} \Bigl( t_1 \leq T^1<t_2, I^1=i, \vX^1 \in B \Bigr) =\\
  \int\limits_{t_1}^{t_2} dt \int\limits_{B}d^2\vx_i \!\! \int\limits_{\prod\limits_{k\neq i} \R_k} \!\! d^{3n-3}x'
  \; \vn_i(\vx_i) \cdot \vj_i^{\psi_t}(x',\vx_i)
\end{multline}
for any $B\subseteq \bou_i$, where $\vj_i$ is the 3-vector that is the component of the current $3n$-vector for particle $i$,
\be
\vj_i^\psi (x) = \frac{\hbar}{m_i} \Im \, \psi^*(x) \nabla_i \psi(x)\,.
\ee
Correspondingly,
\begin{align}
\prob_{\psi_0}(Z^1=\infty) 
&= 1-\sum_{i\in\Lab}\int\limits_{t_0}^{\infty} dt \int\limits_{\bou_i}d^2\vx_i \!\! \int\limits_{\prod\limits_{k\neq i} \R_k} \!\! d^{3n-3}x'
  \; \vn_i(\vx_i) \cdot \vj_i^{\psi_t}(x',\vx_i)\\
  &= \lim_{t\to\infty} \|\psi_t\|^2\,.
\end{align}
In case $Z^1\neq \infty$, and given $T^1,I^1$, and $\vX^1$, restart the problem with the new set of labels $\Lab'=\Lab\setminus\{I^1\}$, the new $3(n-1)$-dimensional configuration space $\R'=\prod_{i\in\Lab'} \R_i$ with elements $x'=(\vx_i)_{i\in\Lab'}$, the new initial time $t_0'=T^1$, the new potential
\be\label{Vprime}
V'(x') = \sum_{i\in\Lab'} V_i(\vx_i) + \frac{1}{2}\sum_{\substack{i,j\in\Lab'\\i\neq j}}V_{ij}(\vx_i,\vx_j)\,,
\ee
and the new $(n-1)$-particle initial wave function
\be\label{newpsi}
\psi_0'(x') = \mathcal{N} \,\psi_{T^1}(x',\vx_{I^1}=\vX^1)
\ee
with normalization constant $\mathcal{N}$. 
Set $T^2=T^{1\prime}, I^2=I^{1\prime}$, and $\vX^2=\vX^{1\prime}$, and repeat the procedure. This completes the statement of the $n$-particle rule.

\subsection{Other Formulations}

Equivalently, the rule can be stated in terms of Bohmian trajectories as follows. At the initial time, choose a Bohmian configuration $X_0$ in $\R$ according to the $|\psi_0|^2$ distribution. Evolve $\psi$ according to the Schr\"odinger equation \eqref{Schrn} with boundary condition \eqref{bcn}, and let the Bohmian configuration $X(t)=\bigl(\vX_1(t),\ldots,\vX_n(t)\bigr)$ move according to Bohm's equation of motion,
\be
\frac{dX(t)}{dt} = \frac{j^{\psi_t}(X(t))}{|\psi_t(X(t))|^2}\,.
\ee
As soon as $X(t)$ reaches $\bou$, i.e., as soon as one of the particles, say particle $i$, reaches its boundary $\bou_i$, set $T^1=t, I^1=i$, and $\vX^1=\vX_i(t)$. Now remove particle $i$ from the configuration while keeping the configuration of the other particles; replace the wave function by the appropriate ``collapsed'' wave function, viz., by \eqref{newpsi}; this kind of wave function, obtained by inserting the actual particle position into some (but not all) of the variables, is well known in Bohmian mechanics as the ``conditional wave function'' \cite{DGZ92,DT,NS14}. 
It follows that the new configuration is $|\psi'|^2$ distributed with respect to the new wave function $\psi'$. Now repeat the procedure. 

Another equivalent formulation of the rule is that
\be
\prob_{\psi_0}\Bigl\{ (Z^1,\ldots,Z^n)\in B \Bigr\} =\scp{\psi_0}{F(B)|\psi_0}
\ee
with POVM $F(\cdot)$ on $\Z$ given by
\begin{align}
&F\Bigl(dt^1\times\{i^1\}\times d^2\vx^1\times \cdots\times dt^n\times\{i^n\}\times d^2\vx^n\Bigr)\nonumber\\
&\qquad= 
L^{1\dagger}\cdots L^{n\dagger} L^n \cdots L^1\; dt^1\, d^2\vx^1\cdots dt^n\, d^2\vx^n\,,\\[3mm]
&L^j\:\:\:\:= \sqrt{\hbar \kappa_{i^j}/m_{i^j}} \; \bigl\langle\vx_{i^j}=\vx^j \bigr|  \,W_{\Lab^{j-1},t^j-t^{j-1}}\,,\label{Ldef}\\[3mm]
&F\Bigl(dt^1\times\{i^1\}\times d^2\vx^1\times \cdots\times dt^r\times\{i^r\}\times d^2\vx^r\times \{\infty\}^{n-r}\Bigr)\nonumber\\
&\qquad= 
L^{1\dagger}\cdots L^{r\dagger} \bigl(\lim_{t\to\infty} W_{\Lab^{r},t}^\dagger W_{\Lab^{r},t}  \bigr) L^r \cdots L^1\; dt^1\, d^2\vx^1\cdots dt^r\, d^2\vx^r\,,
\end{align}
where $\Lab^j=\Lab\setminus \{i^1,\ldots,i^{j}\}$, $\Lab^0=\Lab$, $t^0=t_0$, $W_{\Lab,s}$ is the time evolution according to \eqref{Schrn} and \eqref{bcn} on $\Hilbert_{\Lab}=L^2(\prod_{i\in\Lab}\R_i)$, and $\langle \vx_i=\vx|$ is the mapping from $\Hilbert_{\Lab}$ to $\Hilbert_{\Lab\setminus\{i\}}$ defined by inserting the value $\vx$ for the variable $\vx_i$. The POVM $E(\cdot)$ in \eqref{povm} is given by $E(\cdot)=F(h^{-1}(\cdot))$ where $h$ is the function such that $h(Z^1,\ldots,Z^n)=(Z_1,\ldots,Z_n)$.

\section{Derivation of the $n$-Particle Rule}
\label{sec:derivation}

We give a derivation along the lines of the derivation of the absorbing boundary rule in \cite{detect-rule}; many of the considerations there apply here as well, and we will focus particularly on what is different in the many-particle case. The derivation is particularly easy to understand in the Bohmian picture, although it also works in any other picture of quantum mechanics.

Regard the detectors as a big quantum system $D$ with configuration space $\Q_D=\RRR^{3N}$ (say, $N>10^{23}$), let $P$ denote the $n$-particle system with configuration space $\Q_P=\RRR^{3n}$, and $S=P\cup D$ the composite, with configuration space $\Q_S=\Q_P\times \Q_D$. The $S$ system evolves unitarily with wave function $\Psi$ on $\Q_S$, starting from $\Psi_0=\psi_0\otimes \phi_0$. Let $\Xi_D$ denote the set of $D$-configurations in which the detectors have not clicked but are ready, and $\Upsilon_D$ the set of those in which a detector has fired, so the initial wave function of $D$, $\phi_0$, is concentrated in $\Xi_D$, and $\Psi_0$ is concentrated in $\R\times \Xi_D\subset \Q_S$.

As soon as the first of the $n$ $P$-particles, say particle $i$, reaches its boundary $\bou_i$, the $P$-configuration reaches $\bou$, and the $S$-configuration gets quickly transported from the region $\R\times \Xi_D$ to the region $\Upsilon_S=\Q_P\times \Upsilon_D$. In the interior of $\R\times \Xi_D$, however, no interaction between $P$ and $D$ occurs. 
As discussed in \cite{detect-rule}, the simplest model capturing the right kind of boundary behavior is given by the Schr\"odinger equation with boundary condition $\partial \psi/\partial n = i\kappa \psi$, which in our setting corresponds to \eqref{Schrn} and \eqref{bcn} for all $i\in\Lab$. 

Upon arrival of particle $i$ at $\bou_i$, the $S$-configuration gets moved, more specifically,  to the region $\Upsilon_{S,T_1,I_1,\vX_1}=\Q_P\times \Upsilon_{D,T_1,I_1,\vX_1}$ in which the detector at $\vX_1$ sensitive to particle number $I_1$ is in the triggered state since $T_1$. (For definiteness, the apparatus may be taken to include a clock and a recording device, although the qualitative behavior would presumably be the same without that.) Since the region $\Upsilon_{D,T_1,I_1,\vX_1}$ is macroscopically different from $\Upsilon_{D,t,i,\vx}$ for any other $t$, $i$, or $\vx$, the part of the wave function in that region, which depends on $\psi_0$ only through $\psi_{T_1}(\vx_{I_1}=\vX_1)$, will never again overlap with the parts in other regions $\Upsilon_{S,t,i,\vx}$ or $\R\times \Xi_D$. Thus, conditional on this detection event, the wave function of the configuration $x'$ of the remaining undetected particles is given by \eqref{newpsi}. Now the procedure repeats.

\section{Remarks}
\label{sec:discussion}

\begin{enumerate}
\item\label{rem:moving-several} Suppose now that the boundaries are moving, $\R_i=\R_i(t)$. The rules formulated in Sections~\ref{sec:moving} and \ref{sec:several} can be combined as follows. 
The appropriate boundary condition is 
\be\label{bctn}
\vn_i(t,\vx_i)\cdot \nabla_i \psi(t,x) = i\kappa_i(t,\vx_i)\,\psi(t,x)
\ee
with $\vn_i(t,\vx)$ the outward unit normal vector to $\bou_i(t)$ at $\vx$ and $\kappa_i(t,\vx_i) = \kappa_i + m_i v_{i}(t,\vx_i)/\hbar$, where $v_{i}(t,\vx)$ is the normal velocity (positive, negative, or zero) at which $\bou_i(t)$ moves outward at $\vx$. The Schr\"odinger equation is still \eqref{Schrn}, valid on $\R(t)=\prod_{i\in\Lab}\R_i(t)$. For any $B\subseteq \bigl\{(t,\vx): t>0, \vx\in\bou_i(t)\bigr\}$,
\begin{multline}\label{probnjRnt}
  \prob_{\psi_0} \Bigl( I^1=i, (T^1,\vX^1) \in B \Bigr) =\\
  \int\limits_{B} dt\, d^2\vx_i \!\!\!\! \int\limits_{\prod\limits_{k\neq i} \R_k(t)} \!\!\!\!\!\! d^{3n-3}x'
  \; \Bigl( \vn_i(\vx_i) \cdot \vj_i^{\psi_t}(x',\vx_i) - v_{i}(t,\vx_i) \, |\psi_t(x',\vx_i)|^2 \Bigr)\,,
\end{multline}
and
\be
\prob_{\psi_0} (Z^1=\infty) = \lim_{t\to\infty} \|\psi_t\|^2\,.
\ee
In case $Z^1\neq\infty$, and given $T^1,I^1,\vX^1$, change the wave function according to \eqref{newpsi}, set $\Lab'=\Lab \setminus\{I^1\}$, and repeat the procedure.

\item Suppose now that the particles have spin $\frac{1}{2}$, so $\psi_t:\R\to(\CCC^{2})^{\otimes n}$. Then the rule does not require any change if we understand $\psi^*$ in the formula for the current $\vj$ as the conjugate-\emph{transpose} of $\psi$ (so that $\psi^*\phi$ is the inner product in spin space) and $|\psi|^2$ as $\psi^* \psi$ (i.e., the norm squared in spin space). (However, in the case with spin there is another natural candidate for the formula for $\vj$ besides \eqref{jSchr}, along with a different absorbing boundary condition than \eqref{bc}; see Equation (2) in \cite{detect-dirac}.) We can allow the potential $V$ in the Schr\"odinger equation to be Hermitian-matrix-valued (as it would be in the Pauli equation in the presence of a magnetic field). Note that the wave function $\psi'$ after a detection still takes values in an $n$-particle spin space $(\CCC^{2})^{\otimes n}$ although it depends on the positions of only $n-1$ particles; it is thus different from a usual wave function of $n-1$ particles. That is because, according to the simplest possible concept of an ideal detector, the detector does not interact with the spin degrees of freedom, so that, after a detection, the spin degrees of freedom of the detected particle continue to be entangled with the spin and position degrees of freedom of the other particles, so they cannot be eliminated from the wave function, whereas the position degrees of freedom of the detected particle get disentangled from the other particles (and the spin) by the detection.

In case that after a detection, no interaction occurs any more between the spin of the detected particle and the position or spin degrees of freedom of the other particles, then the spin of the detected particle becomes irrelevant to the joint distribution of $(Z^1,\ldots,Z^n)$, and we can trace it out without affecting that distribution, thus replacing a wave function with $3n-3$ position variables and $n$ spin indices by a density matrix in the Hilbert space of $3n-3$ position variables and $n-1$ spin indices. This replacement would lead to the wrong Bohmian trajectories, but remain empirically equivalent.

An alternative treatment is to equip the detectors with an apparatus that carries out, immediately upon detection of particle $i$ at $(t,\vx)$, a quantum measurement of $\vu\cdot\vsigma$, the component of spin in some direction $\vu=\vu(t,\vx,i)$ (where $\vsigma=(\sigma_x,\sigma_y,\sigma_z)$ is the triple of Pauli spin matrices). Then the outcome of the detection is $Z^j=(T^j,I^j,\vX^j,\Sigma^j)$ with $\Sigma^j=\pm 1$ representing the outcome of the spin measurement, and $\langle \vx_{i^j}=\vx^j|$ in \eqref{Ldef} must be replaced by $\langle \vx_{i^j}=\vx^j| \langle \vu(t^j,\vx^j,i^j)\cdot \vsigma = \Sigma^j|$. The post-detection wave function $\psi'$ then involves the spins (as the positions) of $n-1$ particles.

\item\label{rem:several-imaginary} For a soft detector represented by an imaginary potential \cite{All69}, one can ask the same question we asked in Section~\ref{sec:several}: In a system of several particles that get detected at different times, what is the joint probability distribution of $(Z^1,\ldots,Z^n)$? We mean $Z^j=(T^j,I^j,\vX^j)$ (unless $Z^j=\infty$), but now $\vX^j$ is not limited to a surface but could lie anywhere in the detector volume. 

The natural answer, analogous to our rule of Section~\ref{sec:several}, goes like this: Let $\Gamma_i(t,\vx)\geq 0$ denote the detection rate of particle $i$ at time $t$ and location $\vx\in\RRR^3$.  The wave function $\psi:\RRR^{3n}\to\CCC$ evolves according to
\be\label{SchrIm}
i\hbar\frac{\partial \psi}{\partial t} = -\sum_{i\in\Lab} \frac{\hbar^2}{2m_i} \nabla_i^2 \psi + V\psi - \frac{i\hbar}{2} \sum_{i\in\Lab} \Gamma_i(t,\vx_i) \, \psi
\ee
with potential $V:\RRR^{3n}\to\RRR$ of the form \eqref{V}. The joint distribution of $T^1,I^1,\vX^1$ is given by
\begin{multline}\label{probIm}
  \prob_{\psi_0} \Bigl( t_1 \leq T^1<t_2, I^1=i, \vX^1 \in B \Bigr) =\\
  \int\limits_{t_1}^{t_2} dt \int\limits_{B}d^3\vx_i \!\! \int\limits_{\RRR^{3n-3}} \!\! d^{3n-3}x'
  \; \Gamma_i(t,\vx_i) \, \bigl|\psi_t(x',\vx_i)\bigr|^2
\end{multline}
for any $B\subseteq \RRR^3$. At detection, remove the detected particle and restart the problem with $\Lab'=\Lab\setminus \{I^1\}$, initial time $t_0'=T^1$, new potential $V'$ as in \eqref{Vprime}, and initial $(n-1)$-particle wave function 
\begin{equation}\label{psiprimeIm}
\psi_0'(x') = \mathcal{N}\,\psi_{T^1} \bigl(x',\vx_{I^1}=\vX^1 \bigr)
\end{equation}
with normalization factor $\mathcal{N}$.

Selst\o\ and Kvaal \cite{SK10} considered essentially the same stochastic process for the wave function, although they did not write it down; they wrote down instead the evolution equation of the corresponding density matrix $\rho_t$. They aimed at methods for the numerical simulation of systems of quantum particles in $\RRR^3$ (or unbounded regions), and therefore needed to remove particles that travel away to infinity from the simulation. For our purposes, the density matrix $\rho_t$ would not be useful because it represents an \emph{average} over $Z^1=(T^1,I^1,\vX^1)$ and thus does not allow extracting the \emph{joint} distribution of $Z^1$ and $Z^2$. But for Selst\o\ and Kvaal, $Z^1$ is not a physical event (such as the detection of a particle) but rather the time and place at which a particle was removed from consideration; their interest lay in computing the state of the remaining particles, and for this the average was just appropriate.

In a follow-up work, Selst\o\ \cite{Sel21} developed this scheme further and proposed to expand the part of the $n$-particle wave function lost between $t$ and $t+dt$ into Fourier modes for the removed particle and to collect this information. For our purposes, this option is not available because we ask about the detection times and places, so we need to use the positions basis and not the momentum basis.

Dalibard et al.\ \cite{DCM92} considered a process for $\psi_t$ similar to \eqref{SchrIm}--\eqref{psiprimeIm} but with $\RRR^{3n}$ replaced by a set with 2 elements corresponding physically to an excited state and the ground state of an atom, with Hilbert space $\CCC^2$, with a driving term in the Hamiltonian that will rotate the ground state to the excited state, with $\Gamma=\mathrm{diag}(1,0)$, and with collapse to the ground state---without any reduction in particle number.

\item For other proposed trajectories than Bohm's, such as Deotto--Ghirardi's \cite{DG98} and Nelson's \cite{Nel85,Gol87}, there does not seem to exist any linear boundary condition that would represent ideal detectors that click when and where the particle reaches $\bou$. 
This fits with the fact that those trajectories (in contrast to Bohm's) are often able to move back and forth between rather well-separated wave packets, 
so that the detector cannot be expected to be (irreversibly) triggered when and where the particle arrives; that is, the derivation of the absorbing boundary rule given here (and in \cite{detect-rule} for the single-particle case) would not go through using Deotto--Ghirardi's or Nelson's trajectories. Note also that all three theories (Bohm's, Deotto--Ghirardi's and Nelson's), since they all imply that the configuration of system and apparatus together at any time $t$ is $|\Psi_t|^2$ distributed, make exactly the same empirical predictions, including for detection times, and thus cannot be tested experimentally against each other. The same is true (for the same reason) of different versions of Bohmian mechanics, as long as they preserve the $|\Psi|^2$ distribution.

\item Another absorbing surface was considered in \cite{spacelike}, viz., the space-time singularity $\sS$ of Schwarzschild space-time. Also there, the particle number was decreasing with time, leading to the use of Hilbert spaces with different particle numbers. Of course, no detectors are present along $\sS$, and $\sS$ is spacelike, whereas the boundary in our case should be thought of as timelike. Also, the equations in \cite{spacelike} are based on the Dirac equation, while our discussion here is non-relativistic. However, a deeper difference is that the dynamics of the quantum state in \cite{spacelike} involves the \emph{reduced} quantum state instead of the \emph{conditional} quantum state used here. A reduced quantum state is obtained by tracing out the absorbed degrees of freedom and thus corresponds to taking the \emph{average} over the absorbed degrees of freedom, whereas the conditional quantum state (here, the conditional wave function \eqref{newpsi}) corresponds to \emph{conditioning} on a certain (random) value of the absorbed degrees of freedom. For our problem of detection probabilities, there is no question that conditioning is the correct rule; for one thing, the random value (here, $\vX^1$) gets measured and recorded, and we are interested in the joint distribution of (among other variables) $\vX^1,\ldots,\vX^n$; but even if we ignore the value of $\vX^1$ (and of $T^1$), decoherence caused by the detector will make sure that the coherence is lost between parts of the wave function corresponding to different values of $\vX^1$ (or $T^1$). In contrast, for the question considered in \cite{spacelike}, the behavior at the space-time singularity $\sS$, there seem to be two possibilities for the fundamental Bohmian equation of motion in Schwarzschild space-time, corresponding to using the reduced or the conditional quantum state. The one using the reduced state was spelled out in \cite{spacelike} and leads to a deterministic evolution of a fundamental density matrix $\rho_t$; the other would amount to inserting the (random) space-time location of absorption (on $\sS$) into the $n$-particle quantum state and tracing out the spin of the absorbed particle, leading to a (random) conditional density matrix $\tilde{\rho}_t$ \cite{dm}. In fact, $\rho_t$ is the average of the random $\tilde{\rho}_t$. It seems that the two resulting Bohmian theories are empirically equivalent, and equally natural. While it is not unusual that two different Bohm-type theories are empirically equivalent, it seems to be very rare for them to be equally natural.

\item\label{rem:outside} Suppose now that detected particles do not get removed from the picture but continue to exist and move on outside of $\R$. Then the collapsed wave function $\psi'$ should be, instead of \eqref{newpsi},
\be
\psi_0'(x',\vx_i') = \mathcal{N}' \, \delta^3(\vx_i'-\vX^1)\, \psi_{T_1}(x',\vx_i=\vX^1)
\ee
in case $I^1=i$. That is, the detected particle gets disentangled from the others, and its wave function is a 3-dimensional Dirac delta function; the collapsed wave function $\psi'$ is still a wave function of $n$ particles. The equations governing the further evolution of $\psi'$ in the $\vx_i'$ variable after $T^1$ depend on the physical setup: They may or may not restrict $\vx_i'$ to the complement of $\R_i$ (or to some other region), and they may or may not involve no further detectors capable of detecting particle~$i$ again at a later time.

\item My final remarks are of a more philosophical character. The absorbing boundary rule (for 1 or $n$ particles) arises in a particularly natural way from Bohmian mechanics. Now the Bohmian picture runs against the widespread (and, in my humble opinion, exaggerated) positivistic attitude that in physics one should not talk about the microscopic reality but only about the outcomes of experiments. This attitude often goes with taking the outcomes of experiments as something like the \emph{definition} of reality; for example, this attitude is reflected in calling the time $T$ of detection the ``time of arrival'': What else, asks the positivist, could ``time of arrival'' \emph{mean} than the time at which a detector clicks? According to the positivist, it would be unscientific or even meaningless to talk about an unobserved event such as the arrival of a particle at a certain surface in the absence of detectors there, and if we want to talk about the probability distribution of the arrival time then we must be talking about the distribution of the detection time. However, that is not right; in a universe governed by Bohmian mechanics (and, for all we know, ours might be one), it is entirely meaningful and appropriate to talk about an unobserved arrival. Correspondingly, Bohmian mechanics illustrates how the expression ``time of arrival'' can be ambiguous: Also in the absence of detectors, the particle may arrive at some time $\tau$ at a certain surface, but this time $\tau$ may be different from the time $T$ at which the particle would arrive at the same surface if detectors were present at that surface throughout the experiment; the time $\tau$ has been studied in, e.g., \cite{Daumer,gruebl,KGE03,RGK05,VHD13,DND19}. 

The postivistic attitude may seem to have a trait of modesty (as it would urge us that we claim no more than what we can confirm experimentally), but in practice it can easily lead to the mistake of conflating $T$ and $\tau$. Put differently, unwillingness to consider reality independently of observation can easily make us reason as if the presence of the detector had no back effect on the particle, because it keeps us from making the relevant comparison between the cases with and without detector. The idea that the detector should not have a back effect may have contributed to why the absorbing boundary rule was not discovered earlier and not much appreciated after its discovery. In this rule, after all, the back effect is manifest in that part of the wave function gets reflected from the detecting surface, i.e., that the reflection coefficient is nonzero \cite{detect-rule}.
\end{enumerate}

\end{document}